\documentclass{article}
\usepackage[T1]{fontenc}
\usepackage[utf8]{inputenc}
\usepackage{ismir} 
\usepackage{amsmath,cite,url}
\usepackage{graphicx}
\usepackage{color}
\usepackage{float}
\usepackage{xcolor}

\definecolor{brown}{rgb}{0.59, 0.29, 0.0}
\definecolor{darkgray}{rgb}{0.59, 0.59, 0.59}
\definecolor{tablegray}{gray}{.9}

\newcommand\chris[1]{}
\newcommand\dl[1]{}
\newcommand\aw[1]{}
\newcommand\todo[1]{}

\newcommand\cd[1]{}

\newcommand\rv[1]{\textcolor{black}{#1}}
\newcommand\remove[1]{}

\usepackage{booktabs} 

\usepackage[utf8]{inputenc}
\usepackage{diagbox} 
\usepackage{colortbl}

\usepackage{color}
\usepackage{soul}

\usepackage{tabularx} 

\usepackage{nameref} 

\usepackage{xspace}
\usepackage{enumitem}
\usepackage{mathtools}
\usepackage{commath}

\usepackage{amssymb}
\usepackage{amsmath} 
\usepackage{pifont}

\newcommand{\customtilde}{{\raise.17ex\hbox{$\scriptstyle\sim$}}}

\usepackage{xparse} 
\usepackage{cleveref}

\sethlcolor{yellow}

\title{RISE: Adaptive Music Playback for Realtime Intensity Synchronization with Exercise}

\threeauthors
  {Alexander Wang} {University of Michigan \\ \texttt{alexhci@umich.edu}}
  {Chris Donahue} {Carnegie Mellon University \\ \texttt{chrisdonahue@andrew.cmu.edu}}
  {Dhruv Jain} {University of Michigan \\ \texttt{profdj@umich.edu}}

\def\authorname{A.Wang, C.Donahue, and D.Jain}

\begin{document}

\maketitle
\begin{figure*}
	\centering
	\includegraphics[width=\textwidth]{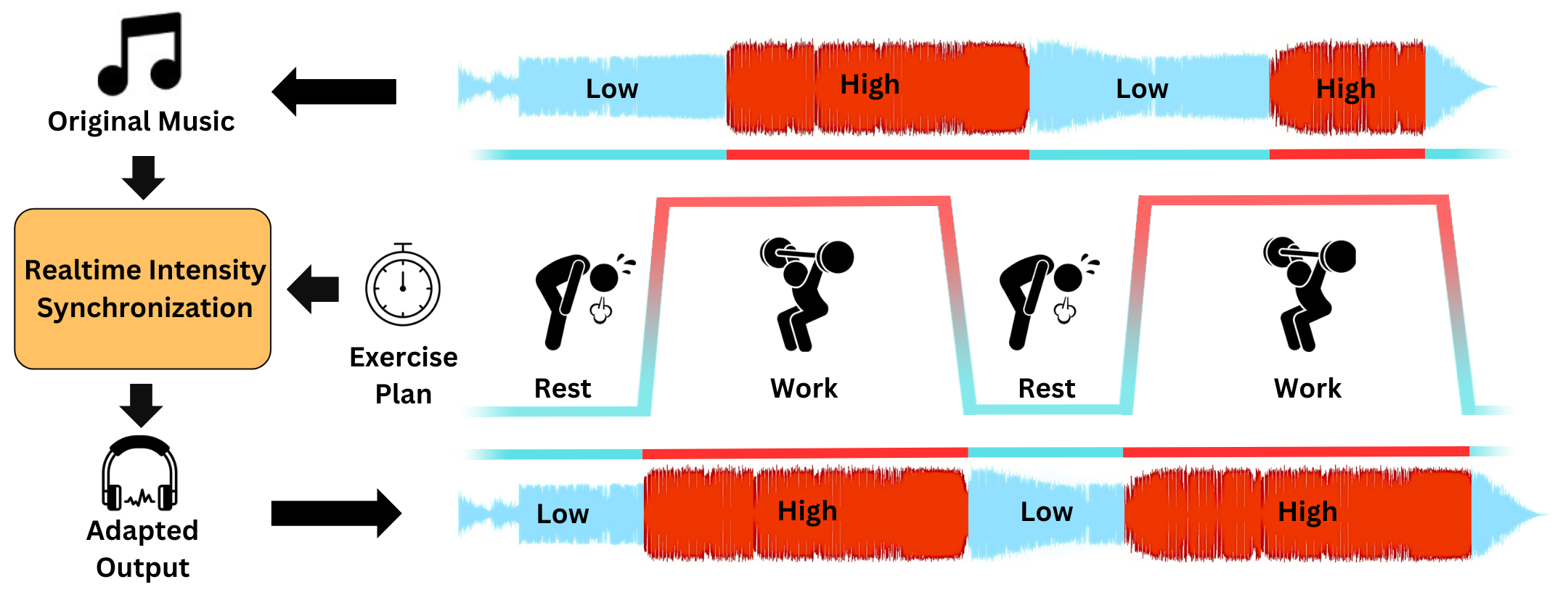}
	\caption{We present RISE, a real-time music intensity synchronization system for exercise. Our system takes in user music and workout phases as input and aligns the high-intensity segments of the music with the user’s exertion phases (and vice versa), to enhance the workout experience. \cd{need to update this figure to change ``rearranged''. also, still worth workshopping if there's a way to more clearly convey the (poor) alignment int he default playback scenario and the (improved) alignment in our playback scenario. right now it requires a bit too much effort IMO to parse the subtle misalignment in the default scenario}}
	\label{fig:teaser}
\end{figure*}
\begin{abstract}

\cd{add youtube link to our system video in a footnote? if you send it to me I can upload to our gclef youtube channel}

We propose a system to adapt a user's music to their exercise by aligning high-energy music segments with intense intervals of the workout. 
Listening to music during exercise can boost motivation and performance. 
However, the structure of the music may be different from the user's natural phases of rest and work, causing users to rest longer than needed while waiting for a motivational section, or lose motivation 
mid-work if the section ends too soon. 
To address this, our system, called RISE, automatically estimates the intense segments in music and uses cutpoint-based music rearrangement techniques to dynamically extend and shorten different segments of the user’s song to fit the ongoing exercise routine. 
\rv{Our system takes as input the rest and work durations to guide adaptation. Currently, this is determined either via a pre-defined plan or manual input during the workout.}\cd{not clear in a vacuum what ``exercise state'' means? need to introduce the term earlier (our system takes as input the state of a user's exercise and ...) or make the sentence interpretable in a vacuum}
We evaluated RISE with 12 participants who compared our system to a non-adaptive music baseline while exercising in our lab. Participants found our rearrangements seamless, intensity estimation accurate, and many recalled moments when intensity alignment helped them push through their workout.


\end{abstract}


\section{Introduction}

The right music can transform an ordinary workout into a more engaging and motivating experience.
Research has shown that listening to music while working out can improve physical performance and reduce perceived exertion \cite{terry2020effects}. Consequently, many commercial applications offer music recommendations to fit the user’s exercise routines \cite{spotifyrun, applefitness}. 
\rv{However, current applications do not consider musical structure, treating songs as uniform in intensity. This can lead to suboptimal scenarios, such as a brief calm section of an otherwise high-energy song playing just as the listener begins weightlifting.}
Conventional music playback remains static and unresponsive to the listener’s dynamic behavior and context, highlighting the need for adaptive music systems that customize the listening experience in real time.


We propose RISE, a real-time intensity synchronization system that \rv{adapts the playback of} user-selected music to better align with exercise plans (\Cref{fig:teaser}). 
\rv{RISE integrates several Music Information Retrieval (MIR) techniques to segment songs, estimate segment intensity, and identify cutpoints, enabling seamless looping and skipping to align segment durations with user exercise.}
RISE operates in two scenarios: in the \emph{guided scenario}, it \rv{adapts} a music recording to fit a predefined workout plan, while in the \emph{unguided scenario}, it adapts in real-time to spontaneous exercise changes, offering flexibility at the potential cost of \rv{adaptation quality}. 
\rv{In this work, we focus on the music analysis and adaptive playback, assuming that the ground truth exercise state is given as input to the system.}

We validate our solution through a two-pronged evaluation. First, we assessed the seamlessness of our system’s \rv{adaptations} by generating excerpts from 270 \remove{Spotify}\cd{not clear what a ``Spotify workout song'' means---is Spotify even necessary to mention at this point?}\remove{workout }songs and conducting an in-lab study in which participants rated transition naturalness. Results showed that our system’s \rv{adaptations} were perceived as seamless, scoring similarly to unmodified excerpts. Second, in a controlled user study, 12 participants compared exercising with \rv{adaptive music} to nonadaptive music. Participants found our \rv{adaptations} seamless and intensity segmentation accurate. They appreciated how the alignment of music with workout enhanced motivation and overall experience, particularly when extending high-energy segments helped them push through. 

In summary, our work makes two primary contributions. 
\rv{First, we propose a novel music adaptation system for exercises by applying several existing MIR techniques.}
\remove{First, we propose a novel music adaption system for exercises using computational music analysis and rearrangement techniques.}
Second, we offer insights from two user studies demonstrating the utility of music \remove{rearrangement}\rv{adaptation} for exercise.
Moreover, as the first to explore \rv{cutpoint-based} rearrangement techniques for adapting music to exercise, 
our work lays the foundation for future research in applying adaptive music systems to scenarios beyond exercise. 
\remove{A supplemental video demonstrating our system is available at: \href{https://youtu.be/XZLBfpt6Lgg}{https://youtu.be/G6gvRIKrGvw.}}
\rv{\footnote{Supplemental video: \url{https://youtu.be/XZLBfpt6Lgg.}}}

\section{Related Work}

Context-aware music systems enrich user experience by dynamically adapting music based on usage contexts, a focus of MIR research emphasizing user-centered applications \cite{knees2019intelligent, Goto2012, schedl2012putting}. For example, prior work has explored adaptive music for enhancing driving experiences \cite{kari2021soundsride, baltrunas2011incarmusic}, notification experiences \cite{Wang24maringba, Wang2024SingingAssistants}, and narrative experiences \cite{shriram2022sonus, 10.1145/2380116.2380163}. Given the well-documented benefits of music in exercise contexts \cite{terry2020effects}, prior work has also explored music adaptations for workouts, including song recommendation \cite{masahiro2008development, elliott2006personalsoundtrack, spotifyrun} and tempo adjustments based on running pace and heart rate \cite{moens2010d, hockman2009real, oliver2006papa, van2011mobeat, chen2024enhancing}. 

While these approaches personalize music at the song level, they overlook the motivational affordances of specific segments within a song. Relaxing music is better suited for recovery, while upbeat, rhythmically prominent music stimulates physical activity \cite{karageorghis2012music}. These variations in intensity also occur within a single song, where high-energy sections like the 'drop' in electronic music feel particularly exciting \cite{turrell2019tension} and exercisers naturally synchronize bursts of effort with these peaks \cite{priest2008qualitative}. We leverage these insights to develop music \rv{adaptation} algorithms that dynamically adjust individual segments within a song, aligning high-intensity music with peak exertion. 

\cd{if you need to save space, probably dont need to mention all the author names inline (citation suffices)}
\rv{RISE combines several MIR techniques for structure-aware adaptive playback. We use the All-in-one package \cite{taejun2023allinone} for \textit{music structure analysis}, automatically segmenting songs into sections (\textit{e.g.,} intro, verse). \cd{spleeter is weirldy personified here relative to its neighbors. ``we use'' vs ``spleeter performs'', also spleeter not introduced by autohr names while others are} Then, we use Spleeter \cite{hennequin2020spleeter} for \textit{source separation}, isolating individual audio components (\textit{e.g.,} drum tracks) to estimate each segment's drum prominence. Finally, we use \textit{cutpoint identification} \cite{plachouras2023music} to pinpoint pairs of transition points where cutting from one to the other wouldn't sound abrupt. This enables RISE to dynamically skip or loop musical segments.}
\remove{Our work is novel in two key ways: (1) it is the first to align music structure with workout structure, adapting within-song intensity to match exercise demands, and (2) it explores seamless cutpoint rearrangement as a technique for real-time adaptation in context-aware music systems, broadening the design space for music adaptation.}
\begin{figure*}[h]
	\centering
	\includegraphics[width=\textwidth]{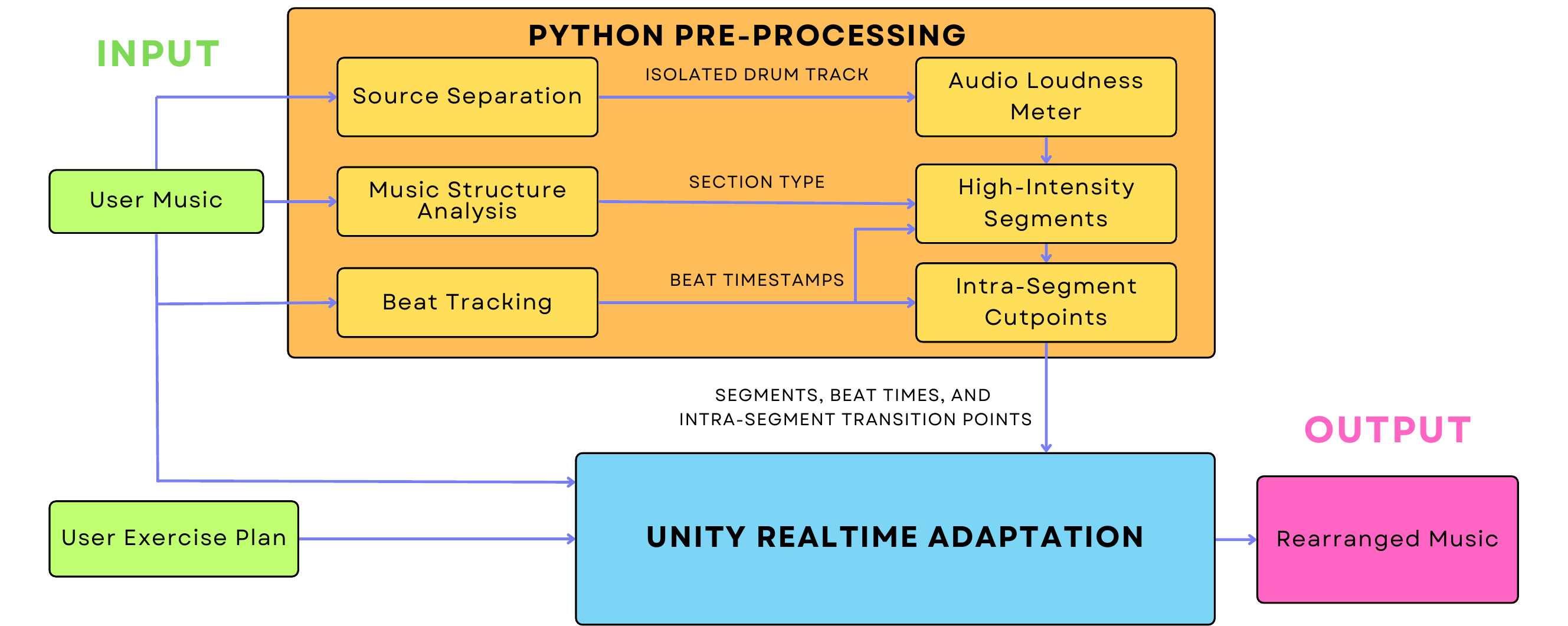}
	\caption{\textbf{System overview.} RISE takes user music and exercise plans as input, preprocesses the music to identify intense segments and intra-segment cutpoints, and sends this information to Unity for real-time adaptation.}
	\label{fig:system}
\end{figure*}
\section{RISE}

We present RISE, a system that synchronizes music intensity with workout intensity to enhance the exercise experience (\Cref{fig:system}). 
RISE takes in a user's music (audio) and workout (planned intervals in the guided setting or current status in the unguided setting) as input. The music is preprocessed to label high-intensity segments (\Cref{sec:pre segment}) and points of transition (\Cref{sec:pre cutpoint}). This information is then used for real-time music \rv{adaptation} by our Unity-based system for workout alignment (\Cref{sec:rearrangement}).
\subsection{Preprocessing - Estimating Intense Segments in Music}\label{sec:pre segment}


\rv{Informed by prior work, we select chorus and instrumental sections as the high-intensity segments, as they are often perceived as more energetic and motivating\cite{turrell2019tension, priest2008qualitative}. To ensure rhythmic drive \cite{LI202424, madison2006experiencing}, we filter out sections lacking in drum presence to avoid arrangements where percussion is intentionally dropped or simplified for contrast.}

We estimate these segments by leveraging two existing music analysis systems. First, we use a music structure analysis system~\cite{taejun2023allinone} to 
segment each song into a set of functional sections 
\( S = \{s_1, s_2, \dots, s_N\} \) and extract beat timestamps 
\( B = \{b_1, b_2, \dots, b_M\} \). 
Here, \( S \) represents 
a partition of the song into $N$ sections, 
where \( s_n \) marks the start timestamp of a section $n$ and \( s_{n+1} \) marks both its end and the start of the next section.
\( B \) is the time-ordered set of detected beat times. Functional section labels (\textit{e.g.,} chorus, instrumental, verse, bridge) are assigned to each section by the system.

\remove{We assume that the perceived intensity of Western pop music is influenced by the prominence of drum elements, which evoke groove and movement \cite{LI202424, madison2006experiencing}.} 
To measure drum prominence, we first extract the drum track using a source separation system~\cite{hennequin2020spleeter}. Given an audio input \( x \), we define the isolated drum signal as \({ d \coloneqq \text{SourceSepDrums}(x) }\).
We then compute LUFS loudness~\cite{steinmetz2021pyloudnorm} at beat-level intervals. For each beat index \( m \), we define its loudness as:  
\begin{equation*}
    L_d(m) \coloneqq \text{LUFS}(\text{Segment}(d, b_m, b_{m+1})),
\end{equation*}
where \(\text{Segment}(d, b_m, b_{m+1})\) extracts the segment of the drum audio between consecutive beats \( b_m \) and \( b_{m+1} \).  

The average drum loudness for a section \( s_n \) is calculated by averaging the loudness of all beats in the segment:  
\begin{equation*}
    L_d(s_n) \coloneqq \frac{1}{|B_n|} \sum_{m \in B_n} L_d(m),
\end{equation*}
where \( B_n \subseteq \{1, \dots, M\} \) is the set of beat indices that satisfy \( s_n \leq b_m < s_{n+1} \).

We precompute a per-song loudness threshold $\tau$, where any segment with a loudness above $\tau$ is considered high intensity, and otherwise we consider it low intensity. Threshold \( \tau \) is computed relative to the loudest section in the song:  
\( \tau = \max_{s_n \in S} L_d(s_n) - \delta \), where \( \delta \) is a constant that determines the relative threshold. In our implementation, we set \( \delta = 5 \) decibels. If four or more consecutive segments are labeled high-intensity, we reassign the segment with the lowest loudness as low-intensity and merge consecutive segments with the same intensity labels. Ultimately, our analysis induces a partition of the full track into $\leq N$ segments and associated binary intensity labels $S' = \{ (s'_1, i_1), (s'_2, i_2), \ldots \}$, where $s'_n$ are delineating timestamps and $i_i \in \{\text{Low}, \text{High}\}$.



\subsection{Preprocessing - Estimating Intra-Segment Cutpoints} \label{sec:pre cutpoint}

To better align intense segments of music with high-intensity workout phases, we estimate a set of \emph{cutpoints} to facilitate seamless \rv{adaptation}. A cutpoint is a pair of timestamps in the music recording, consisting of a starting timestamp and a destination timestamp. The objective is to estimate cutpoints that allow smooth musical transitions, such that if playback jumps from the start of a cutpoint to its destination, users experience minimal disruption.

We derive an initial set of cutpoints using the approach proposed by Plachouras and Miron~\cite{plachouras2023music}, which analyzes recurrence matrices encoding the self-similarity of musical beats. Cutpoints are identified by detecting diagonals in these matrices that correspond to repeated patterns, pinpointing transitions between musically coherent sections. This results in an initial set of candidate cutpoints:
\begin{equation*}
C = \{(c_i^{\text{orig.}}, c_i^{\text{dest.}}) \in B \times B\}
\end{equation*}
where \( C \) is the set of all estimated cutpoint pairs, and each cutpoint consists of a start time \( c_i^{\text{orig.}} \) and an end time \( c_i^{\text{dest.}} \), both aligned to detected beat timestamps \( B \).

In prior work, cutpoints have been applied to rearrange music to fit external constraints, such as video duration~\cite{adoberemix}. These approaches allow cutpoints to cross section boundaries, maximizing flexibility at the potential cost of playback naturalness.

To prioritize naturalness in our system, we enforce an \emph{intra-segment} constraint, ensuring that cutpoints only jump within a segment as opposed to across segments. 
We define the filtered set of intra-segment cutpoints as: 
\begin{equation*}
    C' \coloneqq \{ (c_i^{\text{orig.}}, c_i^{\text{dest.}}) \in C \mid \exists s'_n \in S', \; s'_n \leq c_i^{\text{orig.}}, c_i^{\text{dest.}} < s'_{n+1} \}
\end{equation*}
This guarantees that every cutpoint's start and destination timestamps fall within the same functional section \( s'_n \), preserving the structural integrity of the music. 

\begin{figure}[h]
	\centering
	\includegraphics[width=\linewidth]{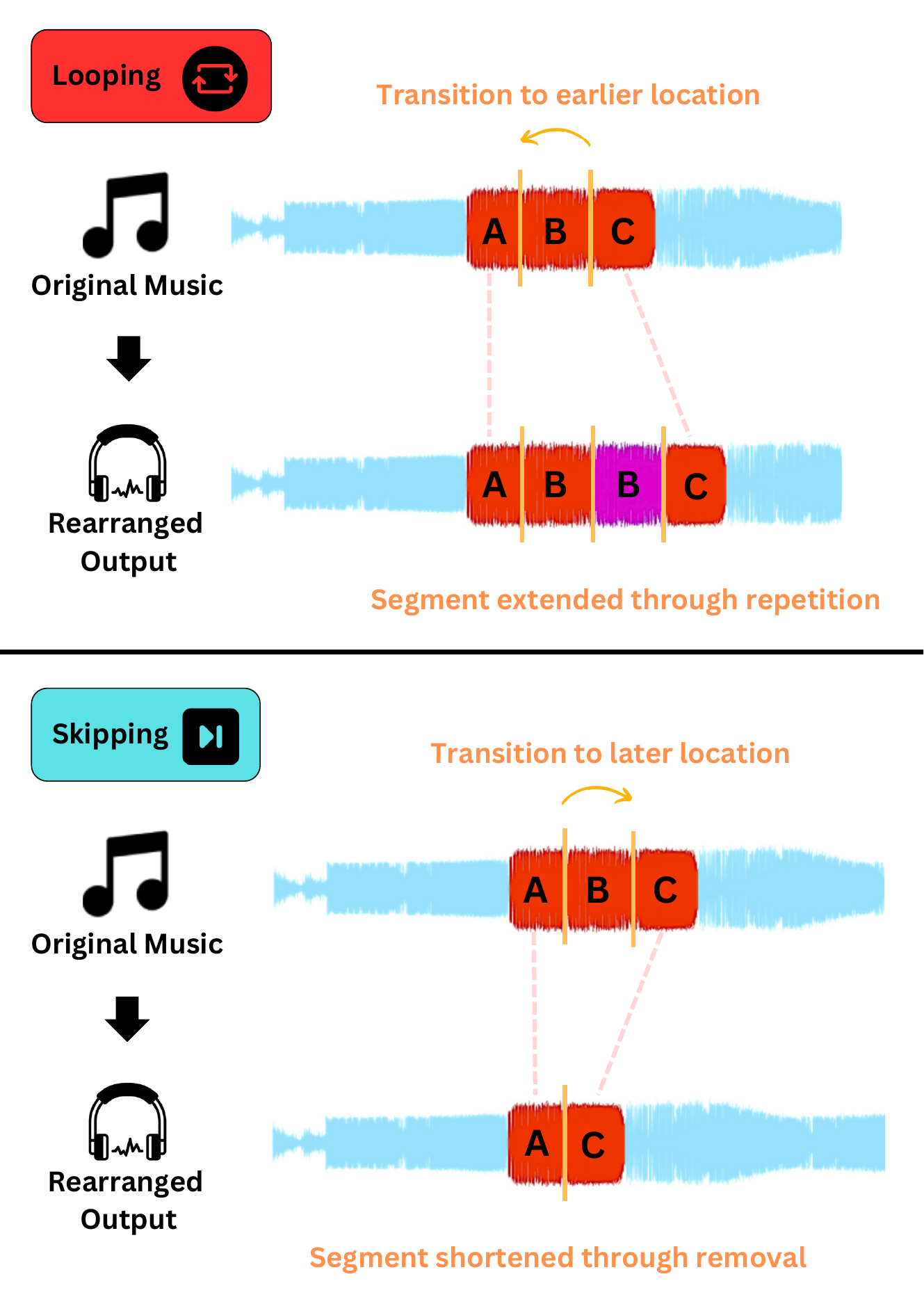}
	\caption{\textbf{Visualization of \rv{adaptation} modes.}
    Top: Loop mode extends a segment by jumping back. Bottom: Skip mode shortens a segment by jumping forward.
    }
	\label{fig:rearrangement}
\end{figure}

\subsection{\rv{Adaptation} With Cutpoints} \label{sec:rearrangement}

In addition to music audio and exercise plan, our real-time \rv{adaptation} system takes as input the following information estimated during pre-processing: 
(1)~musical segments and corresponding intensities,
(2)~seamless cutpoints, and
(3)~beat timestamps. 
To \rv{adapt} music in real time, we define a \textbf{state machine} that governs playback behavior. The system operates in one of three possible states (\Cref{fig:rearrangement}):

\begin{itemize}
    \item \textbf{Loop State}: If the system determines that a segment should be extended, it selects a cutpoint \( c_i \) where \( c_i^{\text{orig.}} > t_{\text{current}} \) and \( c_i^{\text{dest.}} < t_{\text{current}} \), looping previously played sections to increase the duration of the current intensity segment.

    \item \textbf{Skip State}: If a segment duration needs to be shortened, the system selects a cutpoint \( c_i \) where \( c_i^{\text{orig.}} > t_{\text{current}} \) and \( c_i^{\text{dest.}} > c_i^{\text{orig.}} \), skipping forward to reduce the segment duration.

    \item \textbf{Unmodified State}: If no transition is required, the music plays continuously without alteration.
\end{itemize}

Playback state transitions are determined dynamically based on the user’s exercise plan, enabling the system to adjust segment duration in real time.



\subsubsection{Filter-based transitions}
Sometimes, the system cannot immediately transition to the target intensity because there are no available cutpoints that matches the desired transition. 
In these cases, we use a \emph{filter-based transition}, gradually removing high-frequency content before the transition and restoring it afterward, similar to DJ fade techniques. These transitions are noticeable and less seamless compared to cutpoint transitions.
We currently apply filter-based transitions only in unguided mode (\cref{sec:unguided}), when the exercise calls for a timely transition to a high-intensity music state.

\subsection{Usage Modes}

Workout habits may vary greatly across different types of exercises---weight training can require minutes of rest to fully recover and the actual timing can vary greatly depending on how exhausted the user is. Guided interval workouts, on the other hand, emphasize short bursts followed by short rest periods that are strictly timed to maximize time efficiency. Motivated by this observation, we designed two usage modes for two scenarios. An \textit{unguided mode} where the user is free to rest as long as they need, 
and a \textit{guided mode} where the user follows a predefined exercise plan. 
We detail the system design of each mode below.

\subsubsection{Unguided use}
\label{sec:unguided}
\begin{figure}
	\centering
	\includegraphics[width=\linewidth]{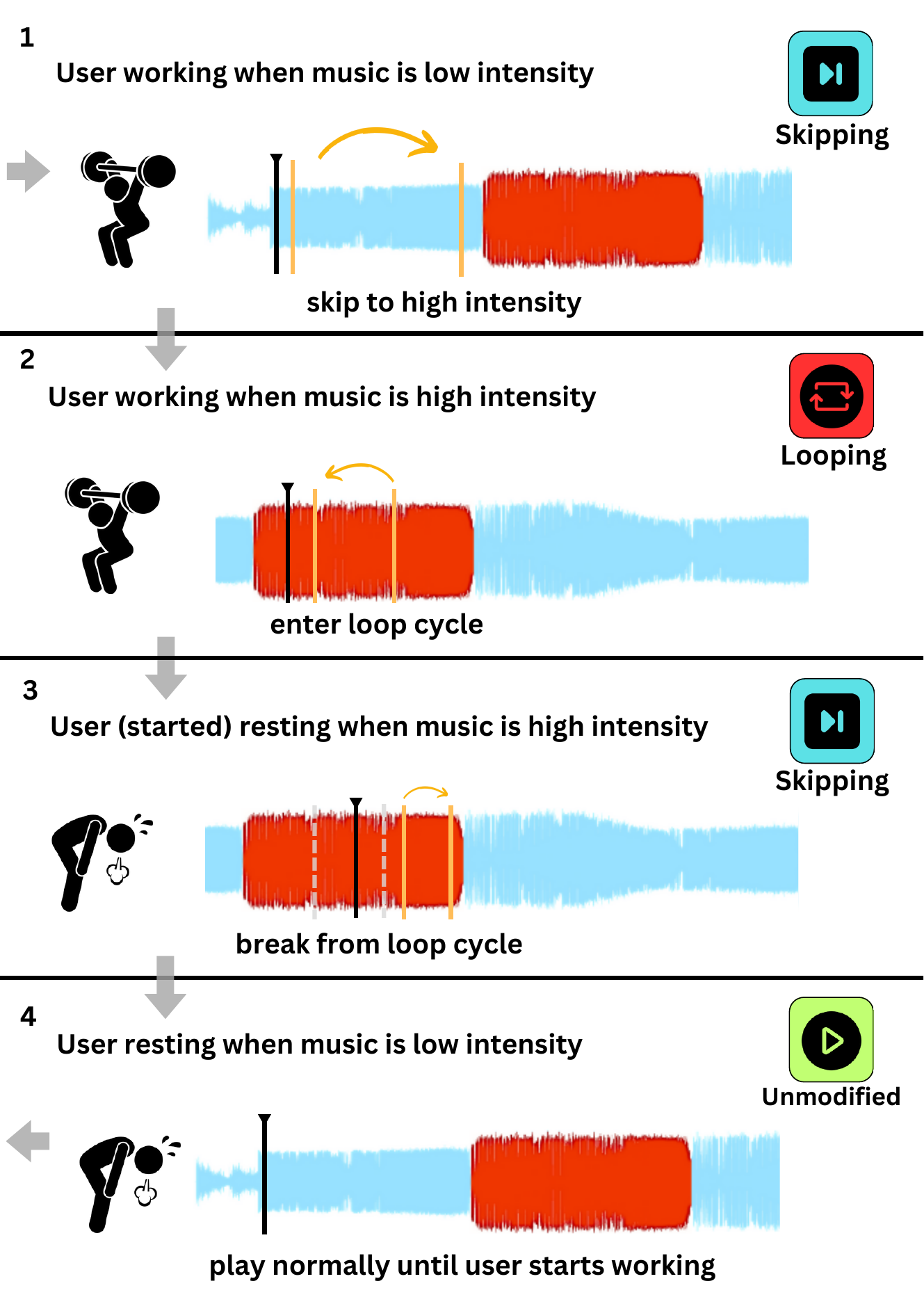}
	\caption{
    Adaptation mode for different scenarios. 
    }
	\label{fig:unguided}
\end{figure}

The unguided mode allows users to freely choose start times and work/rest durations, but sometimes sacrifice \rv{adaptation} quality by using filter-based transitions when no cutpoints are immediately available. The system takes real-time workout state as input and adjusts the music accordingly (binary: work/rest). Currently, users manually indicate state changes by pressing a button.
As depicted in \Cref{fig:unguided}, RISE transitions between playback states depending on the current workout status:
\begin{enumerate}
    \item \textbf{Starting exercise during a low-intensity segment}: The system enters \textbf{skip mode} to quickly transition to a high-intensity segment.  
    \item \textbf{During exercise in a high-intensity segment}: The system activates \textbf{loop mode} to sustain high-intensity music until the user begins resting.  
    \item \textbf{Starting rest in a high-intensity segment}: The system disables looping and switches to \textbf{skip mode} to exit high-intensity segments.  
    \item \textbf{During rest in a low-intensity segment}: The system enters \textbf{unmodified mode}, allowing the music to play naturally until the user resumes exercising.  
\end{enumerate}
\subsubsection{Guided use}
The guided mode provides seamless \rv{adaptations} but requires a precise exercise plan. The system takes two durations (in seconds) for work and rest\remove{ and alternates them when rearranging segments}. We iterate through available cutpoints and select the transition that results in an \rv{adaptation} closest to the desired duration. We allow one transition per segment to maximize naturalness. The results are close to the specified duration but rarely perfect (\textit{e.g.,} adjusting a 50-second segment to 32 seconds for a 30-second target). We fade songs in at the beginning and out at the end, adjusting the intro to match the rest duration.

\section{Quantitative Evaluation}

We conducted an in-lab quantitative listening evaluation to assess the seamlessness of transitions.

\subsection{Procedure}

We collected 270 different songs, 30 songs each from nine different official Spotify workout playlists with different genre preferences.
13 songs were removed from the study because they had similar intensity throughout the entire song or had no available intra-segment cutpoints.
For the remaining 257 songs, we generated two 10-second audio clips per song, one with a transition and one unmodified. Each clip is selected from a random segment of the song. We randomized the transition timing to occur between two and eight seconds within the 10-second clip. 

This evaluation was performed with six human listeners. Each clip was randomly assigned to two listeners who rated transition naturalness on a scale of 1-5 (1 = very jarring, 5 very seamless/unnoticeable). We compare the ratings of modified and unmodified clips using a paired t-test.


\subsection{Results}


We found no statistically significant differences between the clips with transitions (\textit{M =} 4.3/5, \textit{SD =} 1.1) and the baseline clips (\textit{M =} 4.5/5, \textit{SD =} 1.0), \textit{t}(256) = -1.63, \textit{p} = .10, suggesting that our transitions are highly seamless and comparable to the unmodified clips.
We observed a small decrease in the average rating of transition clips (4.5 vs. 4.3) for two reasons. First, the beat detection algorithm is not perfect. We found instances where the transitions were not perfectly aligned, causing a slight jump in rhythm. Second, familiarity with a song influenced the detection of transitions. Raters noted that, even when transitions were completely natural, they perceived differences in expected progression (\textit{e.g.,} altered lyrics) in songs they knew well.
We also found that unmodified clips did not receive perfect scores. This was due to structural elements such as syncopated rhythms and abrupt breaks, intended to surprise the listener, being perceived as transitions by raters, despite these elements being part of the original compositions.



\section{User Study}
We conducted a user study to explore how users experience RISE in both guided and unguided exercise settings.

\subsection{Study Design}

Participants exercised to both unmodified music and our adaptive system across two blocks: guided interval training and unguided weight training. Within each block, they experienced both adaptive and non-adaptive conditions, with order fully counterbalanced. Each condition included a brief tutorial, ~8 minutes of exercise, and optional rest. Participants selected two songs from a curated pool of 25 tracks, played identically across all conditions. The adaptive system modified the music in response to user activity, while the non-adaptive version left the music unchanged. The full session lasted approximately 90 minutes.



\textbf{Guided interval training. }
Participants performed interval exercises of their choice (\textit{e.g.,} jumping jacks) according to a 40s work / 30s rest schedule. 
For the non-adaptive system, these intervals are strict. For the adaptive system, the actual timer may vary by seconds depending on the available transitions in each section. 
Instructions were shown on a screen with countdown visuals and sounds, modeled after popular workout timer videos \cite{intervalTimer}.


\textbf{Unguided weight training.} Participants used dumbbells to perform any freeform weight exercises (\textit{e.g.,} bicep curls). In adaptive conditions, they verbally indicated when they were about to begin or end a work segment, allowing the researcher to input music adaptation state changes. No interface was shown.

\textbf{Interview procedure.} After introducing the study and obtaining participant demographic information and consent, we gave them a verbal description of our system and recorded their first impressions through a short interview. After experiencing all conditions, we showed participants how the system operated by replaying the music for them along with visualizations of both segment intensity labels and cutpoint transitions. This was done at the end of the study to avoid priming participants to focus on specific manipulations and to evaluate whether the \rv{adaptations} were perceptible without guidance. The study concluded with a semi-structured exit interview for verbal feedback.

\textbf{Participant and apparatus.} We recruited 12 participants (5 female, 7 male, age \textit{M} = 24.8 years; \textit{SD} = 2.1) from a local university. Participants regularly exercise (3:~1-2x/week, 7:~3-4x/week, 2:~5-6x/week) for considerable durations (1:~15-30~min, 4:~30-45~min, 1:~45-60~min, 5:~1-2~hours, 1:~2~hours+) in a variety of exercise types (8:~steady cardio, 3:~interval~training, 10:~strength training, 4:~other). The study took place in a controlled lab environment with music played through speakers for participant safety. Participants received \$50 for their participation.


\subsection{Findings}

Interview transcripts were thematically analyzed using a coding reliability approach \cite{boyatzis1998transforming} with the Recal2 tool \cite{recal2}. The analysis resulted in an average raw accuracy of 86.1\% and an average Krippendorff's alpha of 0.72 between two raters, indicating acceptable agreement ($\alpha$ > 0.66). We present key findings below.

\textbf{Participants were excited by the premise of their music adapting to their workouts.} After we explained the study to participants but before they experienced our system, we asked participants about their initial reactions to the description of the study and our proposed system. Some participants had already made manual efforts to align their workout with music (n~=~8), such as coordinating specific music sections with their workout (n~=~4). All participants expressed that music intensity alignment with workouts could be helpful to them. Participants believed alignment can increase motivation, or help them feel more energized and “pumped” during the workout (n~=~11).



\textbf{Estimated high-intensity segments aligned with their intuitions.}
We showed participants a visualization of our system's estimated high-intensity music segments after the study, and 
all participants found the results to be aligned with their expectations (n~=~12). 

\textbf{Participants found our cutpoint \rv{adaptation} technique seamless, subverting their expectations.} Before the study, we shared a high-level description of our workout adaptation system with participants. Many expressed initial skepticism that the modifications might detract from the naturalness of the music playback (n~=~6). 
Out of the participants who had concerns regarding the naturalness of music, all but one remarked that their concerns were overturned after experiencing our system (n~=~5).
While the filter-based modifications were regarded as more noticeable, most participants either did not notice or barely noticed any cutpoint-based modifications until they were shown a visualization at the end of the study (n~=~10). 

\textbf{Participants appreciated the extra motivation our alignment approach provided.} All but one participant found the intensity alignment to enhance their workout experience (n = 11), giving them more motivation to push harder in their workouts. P2 noted that when the music intensity peaked just as they were about to reach failure in their workout, it helped them push through and get an extra one or two reps. Similarly, P4 recalled an instance where they expected the music to shift to a lower-intensity segment in the middle of their work interval, but instead the upbeat segment repeated, providing them the energy to push through the remainder of their workout.

\textbf{Participants varied in their sensitivity to both music modifications and alignment with exercise. }For instance, P8 barely noticed music changes but felt their movements were more consistent in the adaptive condition, while the nonadaptive one felt chaotic. In contrast, P3, who regularly coordinates music with workouts, immediately noticed misalignments in the nonadaptive condition and even felt the urge to stop the music when high-energy segments played during their rest.
Sensitivity could also depend on exercise. 
P6 mentioned that they treated music as background during simple exercises but valued music alignment for motivation during more challenging exercises, such as the bench press.
This indicates the need to better understand how different modes of music engagement influence tolerance and demand for system-driven changes.


\textbf{Participants preferred adaptive music but identified issues with the unguided experience.}
Out of 12 participants, 11 preferred our adaptive system over nonadaptive music: 6 preferred it in both scenarios, while 5 favored it only in the guided setting. Those who preferred the system only in the guided setting found two main issues with the unguided experience. First, the input method—requiring them to notify the researcher before exertion—was distracting and impractical (n = 7). Second, while cutpoint transitions were deemed seamless and often unnoticeable, filter transitions disrupted the natural flow of the music. These filter transitions were especially common during exercises with longer work durations and shorter rest periods, where excessive looping also occurred, leading to unnatural \rv{adaptations} of the music. As a result, participants recommended improvements in system customizability to support different workout habits (\textit{e.g.,} long work, short rest) and music adaptation preferences (n = 8), manipulation seamlessness (n = 4), and expanding the available music pool or allowing user input (n = 3).

\section{Limitation and Future Work}
\remove{Our study revealed two key limitations: first, the need for user input, which could be impractical when holding weights, and second, disruptions caused by filter transitions, which are noticeable transitions used when seamless transitions aren't immediately available on demand. To address limitations, future work could focus on two areas:}
\rv{Our study revealed several limitations, including users finding manual input impractical and filter transitions disrupting. We propose future work below:}

\textbf{Fully automating the system with sensing technologies.} To eliminate the need for manual interaction during workouts, we propose automating the system using real-time sensing technologies. By integrating activity recognition, the system could automatically detect exercise phases without manual input. The system could also model user behavior and predict upcoming exercise states based on past patterns to prepare \rv{adaptation} plans in advance. 

\textbf{Incorporating other types of modifications.} While cutpoints enable seamless transitions, they may be unavailable when timely adaptation is required. Future work could complement our segment-sensitive approach with existing techniques, such as pace adjustment and song recommendation, to compensate for minor time discrepancies. Additionally, exploring audio inpainting with generative models could enable smooth transitions between arbitrary segments, providing greater flexibility and precision.

\rv{\textbf{Determining high-intensity segments.} RISE currently aligns drum-prominent chorus and instrumental sections with user exertion. While this approach worked well in out study, future work can further explore how segment-level musical variations influence exercise and how these effects may differ by genre or listener preference. }

\section{Conclusion}
We present RISE, a novel system that \rv{adapts} music to align with exercise phases. A user study involving 12 participants revealed that, despite initial skepticism, most users appreciated the alignment and preferred it over nonadaptive music for their workouts. Our work represents a step towards expanding the design space of adaptive music, making tailored music experiences, once limited to video games and precomposed soundtracks, applicable to real-world scenarios like workouts.

\clearpage
\bibliography{ISMIRtemplate}

\begin{thebibliography}{10}
\providecommand{\url}[1]{#1}
\csname url@samestyle\endcsname
\providecommand{\newblock}{\relax}
\providecommand{\bibinfo}[2]{#2}
\providecommand{\BIBentrySTDinterwordspacing}{\spaceskip=0pt\relax}
\providecommand{\BIBentryALTinterwordstretchfactor}{4}
\providecommand{\BIBentryALTinterwordspacing}{\spaceskip=\fontdimen2\font plus
\BIBentryALTinterwordstretchfactor\fontdimen3\font minus \fontdimen4\font\relax}
\providecommand{\BIBforeignlanguage}[2]{{%
\expandafter\ifx\csname l@#1\endcsname\relax
\typeout{** WARNING: IEEEtran.bst: No hyphenation pattern has been}%
\typeout{** loaded for the language `#1'. Using the pattern for}%
\typeout{** the default language instead.}%
\else
\language=\csname l@#1\endcsname
\fi
#2}}
\providecommand{\BIBdecl}{\relax}
\BIBdecl

\bibitem{terry2020effects}
P.~C. Terry, C.~I. Karageorghis, M.~L. Curran, O.~V. Martin, and R.~L. Parsons-Smith, ``Effects of music in exercise and sport: A meta-analytic review.'' \emph{Psychological bulletin}, vol. 146, no.~2, p.~91, 2020.

\bibitem{spotifyrun}
S.~Mitroff, ``Hitting the pavement with spotify running (hands-on) - cnet,'' \url{https://www.cnet.com/tech/services-and-software/hitting-the-pavement-with-spotify-running-hands-on/}, 2015.

\bibitem{applefitness}
Apple, ``Apple fitness+ - apple,'' \url{https://www.apple.com/apple-fitness-plus/}, 2024.

\bibitem{knees2019intelligent}
P.~Knees, M.~Schedl, and M.~Goto, ``Intelligent user interfaces for music discovery: The past 20 years and what's to come.'' in \emph{Proceedings of the 20th International Society for Music Information Retrieval Conference, {ISMIR} 2019, Delft, The Netherlands, November 4-8, 2019}, 2019, pp. 44--53.

\bibitem{Goto2012}
M.~Goto, ``Grand challenges in music information research,'' in \emph{Dagstuhl Follow-Ups: Multimodal Music Processing}, M.~M{\"u}ller, M.~Goto, and M.~Schedl, Eds.\hskip 1em plus 0.5em minus 0.4em\relax Dagstuhl Publishing, 2012, pp. 217--225.

\bibitem{schedl2012putting}
M.~Schedl and A.~Flexer, ``Putting the user in the center of music information retrieval.'' in \emph{Proceedings of the 13th International Society for Music Information Retrieval Conference, {ISMIR} 2012, Mosteiro S.Bento Da Vit{\'{o}}ria, Porto, Portugal, October 8-12, 2012}, 2012, pp. 385--390.

\bibitem{kari2021soundsride}
M.~Kari, T.~Grosse-Puppendahl, A.~Jagaciak, D.~Bethge, R.~Sch{\"u}tte, and C.~Holz, ``Soundsride: Affordance-synchronized music mixing for in-car audio augmented reality,'' in \emph{The 34th Annual ACM Symposium on User Interface Software and Technology}, 2021, pp. 118--133.

\bibitem{baltrunas2011incarmusic}
L.~Baltrunas, M.~Kaminskas, B.~Ludwig, O.~Moling, F.~Ricci, A.~Aydin, K.-H. L{\"u}ke, and R.~Schwaiger, ``Incarmusic: Context-aware music recommendations in a car,'' in \emph{E-Commerce and Web Technologies: 12th International Conference, EC-Web 2011, Toulouse, France, August 30-September 1, 2011. Proceedings 12}.\hskip 1em plus 0.5em minus 0.4em\relax Springer, 2011, pp. 89--100.

\bibitem{Wang24maringba}
A.~Wang, Y.~F. Cheng, and D.~Lindlbauer, ``Maringba: Music-adaptive ringtones for blended audio notification delivery,'' in \emph{Proceedings of the {CHI} Conference on Human Factors in Computing Systems, {CHI} 2024, Honolulu, HI, USA, May 11-16, 2024}.\hskip 1em plus 0.5em minus 0.4em\relax New York, NY, USA: Association for Computing Machinery, 2024.

\bibitem{Wang2024SingingAssistants}
A.~Wang, D.~Lindlbauer, and C.~Donahue, ``Towards music-aware virtual assistants,'' in \emph{Proceedings of the 37th Annual {ACM} Symposium on User Interface Software and Technology, {UIST} 2024, Pittsburgh, PA, USA, October 13-16, 2024}.\hskip 1em plus 0.5em minus 0.4em\relax New York, NY, USA: Association for Computing Machinery, 2024.

\bibitem{shriram2022sonus}
J.~Shriram, M.~Tapaswi, and V.~Alluri, ``Sonus texere! automated dense soundtrack construction for books using movie adaptations,'' in \emph{Proceedings of the 23rd International Society for Music Information Retrieval Conference, {ISMIR} 2022, Bengaluru, India, December 4-8, 2022}, 2022, pp. 535--542.

\bibitem{10.1145/2380116.2380163}
S.~Rubin, F.~Berthouzoz, G.~Mysore, W.~Li, and M.~Agrawala, ``Underscore: musical underlays for audio stories,'' in \emph{Proceedings of the 25th Annual ACM Symposium on User Interface Software and Technology}, ser. UIST '12.\hskip 1em plus 0.5em minus 0.4em\relax New York, NY, USA: Association for Computing Machinery, 2012, p. 359–366.

\bibitem{masahiro2008development}
N.~Masahiro, H.~Takaesu, H.~Demachi, M.~Oono, and H.~Saito, ``Development of an automatic music selection system based on runner’s step frequency,'' in \emph{{ISMIR} 2008, 9th International Conference on Music Information Retrieval, Drexel University, Philadelphia, PA, USA, September 14-18, 2008}, 2008, pp. 193--198.

\bibitem{elliott2006personalsoundtrack}
G.~T. Elliott and B.~Tomlinson, ``Personalsoundtrack: context-aware playlists that adapt to user pace,'' in \emph{CHI'06 extended abstracts on Human factors in computing systems}, 2006, pp. 736--741.

\bibitem{moens2010d}
B.~Moens, L.~van Noorden, and M.~Leman, ``D-jogger: Syncing music with walking,'' in \emph{7th Sound and music computing Conference}.\hskip 1em plus 0.5em minus 0.4em\relax Universidad Pompeu Fabra, 2010, pp. 451--456.

\bibitem{hockman2009real}
J.~Hockman, M.~M. Wanderley, and I.~Fujinaga, ``Real-time phase vocoder manipulation by runner's pace.'' in \emph{NIME}, 2009, pp. 90--93.

\bibitem{oliver2006papa}
N.~Oliver and L.~Kreger-Stickles, ``Papa: Physiology and purpose-aware automatic playlist generation.'' in \emph{ISMIR 2006, 7th International Conference on Music Information Retrieval}, 2006, pp. 250--253.

\bibitem{van2011mobeat}
B.~van~der Vlist, C.~Bartneck, and S.~M{\"a}ueler, ``mobeat: Using interactive music to guide and motivate users during aerobic exercising,'' \emph{Applied psychophysiology and biofeedback}, vol.~36, pp. 135--145, 2011.

\bibitem{chen2024enhancing}
Y.~Chen, C.-C. Chen, L.-C. Tang, and W.-H. Chieng, ``Enhancing running exercise with iot, blockchain, and heart rate adaptive running music,'' \emph{IEEE Access}, 2024.

\bibitem{karageorghis2012music}
C.~I. Karageorghis and D.~Holland, ``Music in the exercise domain: A review and synthesis (part ii),'' \emph{International Review of Sport and Exercise Psychology}, vol.~5, no.~1, pp. 67--84, 2012.

\bibitem{turrell2019tension}
A.~Turrell, A.~R. Halpern, and A.-H. Javadi, ``When tension is exciting: an electroencephalogram exploration of excitement in music,'' \emph{bioRxiv}, p. 637983, 2019.

\bibitem{priest2008qualitative}
D.-L. Priest and C.~I. Karageorghis, ``A qualitative investigation into the characteristics and effects of music accompanying exercise,'' \emph{European physical education review}, vol.~14, no.~3, pp. 347--366, 2008.

\bibitem{taejun2023allinone}
T.~Kim and J.~Nam, ``All-in-one metrical and functional structure analysis with neighborhood attentions on demixed audio,'' in \emph{IEEE Workshop on Applications of Signal Processing to Audio and Acoustics (WASPAA)}, 2023.

\bibitem{hennequin2020spleeter}
R.~Hennequin, A.~Khlif, F.~Voituret, and M.~Moussallam, ``Spleeter: a fast and efficient music source separation tool with pre-trained models,'' \emph{Journal of Open Source Software}, vol.~5, no.~50, p. 2154, 2020.

\bibitem{plachouras2023music}
C.~Plachouras and M.~Miron, ``Music rearrangement using hierarchical segmentation,'' in \emph{ICASSP 2023-2023 IEEE International Conference on Acoustics, Speech and Signal Processing (ICASSP)}.\hskip 1em plus 0.5em minus 0.4em\relax IEEE, 2023, pp. 1--5.

\bibitem{LI202424}
C.-W. Li and C.-G. Tsai, ``The presence of drum and bass modulates responses in the auditory dorsal pathway and mirror-related regions to pop songs,'' \emph{Neuroscience}, vol. 562, pp. 24--32, 2024.

\bibitem{madison2006experiencing}
G.~Madison, ``Experiencing groove induced by music: consistency and phenomenology,'' \emph{Music perception}, vol.~24, no.~2, pp. 201--208, 2006.

\bibitem{steinmetz2021pyloudnorm}
C.~J. Steinmetz and J.~D. Reiss, ``pyloudnorm: {A} simple yet flexible loudness meter in python,'' in \emph{150th AES Convention}, 2021.

\bibitem{adoberemix}
Adobe, ``Remix in premiere pro,'' \url{https://helpx.adobe.com/premiere-pro/using/remix-audio-in-premiere-pro.html}, 2024.

\bibitem{intervalTimer}
\BIBentryALTinterwordspacing
W.~M.~W. Timer, ``Interval timer with music | 40 sec rounds 30 sec rest | mix 107,'' YouTube video, 2021, accessed: September 1, 2024. [Online]. Available: \url{https://www.youtube.com/watch?v=lnBOQnc_p-E}
\BIBentrySTDinterwordspacing

\bibitem{boyatzis1998transforming}
R.~E. Boyatzis, \emph{Transforming qualitative information: Thematic analysis and code development}.\hskip 1em plus 0.5em minus 0.4em\relax Sage, 1998.

\bibitem{recal2}
D.~Freelon, ``Recal2: Reliability for 2 coders,'' \url{http://dfreelon.org/utils/recalfront/recal2/}, 2010.

\end{thebibliography}

%
%
%
%

\end{document}